\documentclass[english]{article}
\date{}
\usepackage{authblk}
\usepackage{pgfplots}
\usepackage{amsmath,amsthm,amssymb}
\usepackage{tikz}
\usepackage{pgf-pie}
\usetikzlibrary{arrows,automata}
\usepackage{caption}
\usepackage{subcaption}
\usepackage{makecell}
\usepackage{float}
\usetikzlibrary{shapes,arrows,automata}
\usepackage{dot2texi}
\usepackage[boxed,linesnumbered]{algorithm2e}
\usepackage{hyperref}
\hypersetup{
colorlinks=true,
citecolor=blue
}

\begin{document}
\title{A Note on the Ranking of Saudi Arabian Universities based on highlycited.com}
\author[1,2]{Eisa Alanazi\thanks{\texttt{E-Mail: alanazie@cs.uregina.ca}}}

\affil[1]{\small Department of Computer Science, University of Regina, Regina, Canada}
\affil[2]{\small Department of Computer Science, Umm AlQura University, Makkah, Saudi Arabia}

\renewcommand\Authands{ and }
\maketitle

\begin{abstract}
Recently, Thomson Reuters has published its 2014 list of highly cited researchers (HCRs)\cite{HCR}. Initial studies over the list \cite{DBLP:journals/corr/BornmannB14} suggested that some universities (for instance, King Abdulaziz University in Saudi Arabia) may have been manipulating its world ranking by contracting with highly cited researchers. In this work, we analyse the ranking of other Saudi universities based solely on the list. Our analysis suggests that other universities in Saudi Arabia do not follow the steps of King Abdulaziz University when it comes to contracting with HCRs. 
\end{abstract}

\section{Preprocessing the data }
Before analysing the 2014 HCRs list, we need to preprocess the  data for the following main reasons: 
\begin{itemize}
\item Inconsistency on institutes names. 
\item Handling duplicates on the list.  
\end{itemize}

For inconsistent naming, we make sure that the affiliation name is the same across the list. For instance, King Abduluziz University is the same as King Abdulaziz University or Urbana Champaign and Urbana-Champaign. There are other minor inconsistencies including missing the country of some institutes while mention it in other entries.  For duplicate authors, there are two researchers with different primary affiliations across the list and one author with different secondary affiliation.

In this work, we do not try to combine different institutions into a single one. Instead, we count them as if they were different even though, in reality, they may belong to the same institution. Thus, we do not combine all universities belong to the University of California system into one institute. 

We see the recent list more informative than the former one (i.e., the 2001 list) in a sense that it makes a clear distinction between primary and secondary affiliations  for a given researcher. 
\section{Institutes Ranking} 
\begin{table}[H]
\centering
\begin{tabular}{|c|c|c|}
\hline Rank&Institute Name& Number\\
\hline 1& Harvard University & 100 \\
\hline 2& Stanford University & 51 \\
\hline 3& Chinese Academy of Sciences & 41 \\
\hline 4& University of California, Berkeley& 38 \\
\hline 5& NIH & 32 \\
\hline 6& University of Oxford & 31 \\
\hline 7& Duke University & 29 \\
\hline 8& Massachusetts Institute of Technology (MIT)& 28\\
\hline 9& University of California, San Diego& 27 \\
\hline 9& University of Michigan - Ann Arbor & 27 \\
\hline 9& Northwestern University & 27\\
\hline 9& Princeton University & 27 \\
\hline 10& University of Washington& 26\\
\hline 11& Brigham \& Womens Hosp& 24\\
\hline 12& University of California, Los Angeles& 23 \\
\hline 12& Wellcome Trust Sanger Inst& 23\\
\hline 13& University of Cambridge &22\\
\hline 13& University of California, Santa Cruz& 22\\
\hline 14& Cornell University & 21\\
\hline 14& Imperial College London& 21\\
\hline 14& The Johns Hopkins University & 21\\
\hline 14& University of Chicago& 21\\
\hline 14& Columbia University& 21\\
\hline 15& Mayo Medical School & 20\\
\hline
\end{tabular}
\caption{Institutes ranking based on primary affiliations}
\end{table}

\begin{table}[H]
\centering
\begin{tabular}{|c|c|c|}
\hline Rank&Institute Name& Number\\
\hline 1& King Abdulaziz University & 133\\
\hline 2& Harvard University & 35 \\
\hline 3& Massachusetts Gen Hosp& 11\\
\hline 4& Massachusetts Institute of Technology (MIT)& 10\\
\hline 5& The University of Tokyo& 9\\
\hline 6& University of Copenhagen&8\\
\hline 7& University of California, Berkeley& 7\\
\hline 7& University of Melbourne& 7\\
\hline 8& University of Cambridge & 6\\
\hline 9&Howard Hughes Med Inst& 5\\
\hline 9&Dana Farber Canc Inst& 5\\
\hline 9&Inserm& 5\\
\hline 9&Imperial College London& 5\\
\hline 10& Stanford University & 4\\
\hline 10& Monash University & 4\\
\hline 10& University of Chicago & 4\\
\hline 10& University of California, San Francisco & 4\\
\hline 10& University of British Columbia & 4\\
\hline 10& New York University & 4\\
\hline 10& University of Toronto & 4\\
\hline 10& University of Oxford & 4\\
\hline 10& Utrecht University & 4\\
\hline 10& University of Wageningen& 4\\
\hline 10& University of Washington & 4\\
\hline 10& King's College London & 4\\
\hline
\end{tabular}
\caption{Institutes ranking based on secondary affiliations}
\end{table}

\begin{table}[H]
\centering
\begin{tabular}{|c|c|c|}
\hline Rank&Institute Name& Number\\
\hline 1& King Abdulaziz University & 144\\
\hline 2& Harvard University & 135 \\
\hline 3& Stanford University & 55\\
\hline 4& University of California, Berkeley& 45\\
\hline 5& Chinese Academy of Sciences & 43 \\
\hline 6& Massachusetts Institute of Technology (MIT)& 38 \\
\hline 7& University of Oxford & 35 \\
\hline 8& NIH & 34 \\
\hline 9& University of Washington& 30\\
\hline 10& Princeton University & 29\\
\hline 10& Duke University & 29 \\
\hline 11& University of Cambridge & 28\\
\hline 11& University of Michigan - Ann Arbor & 28\\
\hline 11& Northwestern University & 28\\
\hline 12& University of California, San Diego& 27\\
\hline 13& Wellcome Trust Sanger Inst& 26 \\
\hline 13& Imperial College London & 26\\
\hline 14& Brigham \& Womens Hosp & 25\\
\hline 14& University of Chicago & 25\\
\hline 15& University of California, Los Angeles& 24\\
\hline 15& The Johns Hoplins University & 24\\
\hline 15& The University of Tokyo& 24\\
\hline
\end{tabular}
\caption{Institutes ranking based on primary and secondary affiliations}
\end{table}

\subsection{Country Ranking}
\begin{figure}[H]
\begin{tikzpicture}[scale=.87]
\pie[text=legend]{61/USA, 11.2/UK, 5.8/Germany, 5.3/Saudi Arabia, 5.3/China, 11.4/others}
 \end{tikzpicture}
 \caption{HCRs country distribution based on primary and secondary affiliations} 
 \end{figure}
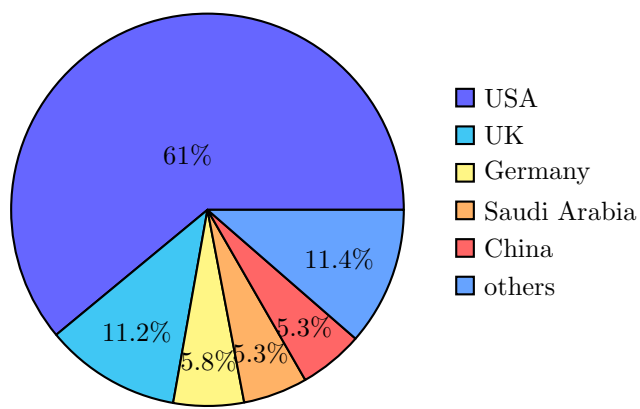
 
\section{Saudi Universities}
There are four Saudi universities mentioned on the list: King Abdulaziz University (KAU), King Saud University (KSU), King Abdullah University of Science and Technology (KAUST) and King Fahd University of Petroleum and Minerals (KFUPM). In Table \ref{sa_ranking}, we list their ranking based on either primary or secondary affiliations. For example, KAU is ranked 24${th}$ when considering the number of HCRs primary affiliated with it while its ranked the 1${st}$ when it comes to the number of HCRs mentioned it as a secondary affiliation. 

\begin{table}[H]
\centering
\begin{tabular}{|c|c|c|c|}
\hline Name & Primary (out of 34) & Secondary (out of 13) & Both (out of 38)\\
\hline KAU       & 24  & 1  & 1\\
\hline KSU       &   23&   12 & 25\\
\hline KAUST  &   31&    N/A& 35\\
\hline KFUPM  &   34&   13 & 37\\

\hline 
\end{tabular}
\caption{Saudi universities ranking based on primary/secondary affiliations}
\label{sa_ranking}
\end{table}

Similarly, Table \ref{sa_number} shows, for every Saudi university, the number of primary affiliated HCRs and the number of secondary affiliated HCRs. As mentioned in \cite{DBLP:journals/corr/BornmannB14}, King Abdulaziz University (KAU) has a weird standing with this regard. While there are 11 researchers primary affiliated with it, there are 133 researchers mentioned it as their secondary affiliation. Other universities have a reasonable trade-off between these two numbers. 

\begin{table}[H]
\centering
\begin{tabular}{|c|c|c|c|}
\hline Name & \#Primary & \#Secondary & Total\\
\hline KAU       & 11  & 133   &144 \\
\hline KSU       &   12&   2 & 14\\
\hline KAUST  &   4&    0&4\\
\hline KFUPM  &   1&   1&2\\

\hline 
\end{tabular}
\caption{Number of HCRs affiliated with Saudi universities. }
\label{sa_number}
\end{table}

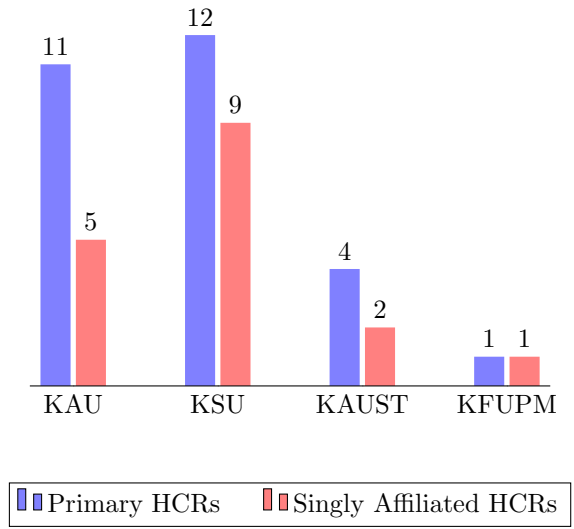
\begin{figure}[H]
\centering
\begin{tikzpicture}
  \centering
  \begin{axis}[
        ybar, axis on top,
        height=8cm, width=8.5cm,
        bar width=0.4cm,
        ymajorgrids, tick align=inside,
        major grid style={draw=white},
        enlarge y limits={value=.1,upper},
        ymin=0, ymax=15,
        axis x line*=bottom,
        axis y line*=right,
        y axis line style={opacity=0},
        tickwidth=0pt,
        ytick=\empty,
        enlarge x limits=true,
        legend style={
            at={(0.5,-0.2)},
            anchor=north,
            legend columns=-1,
            /tikz/every even column/.append style={column sep=0.5cm}
        },
        symbolic x coords={
           KAU,KSU,KAUST,KFUPM
          },
       xtick=data,
       nodes near coords={
        \pgfmathprintnumber[precision=0]{\pgfplotspointmeta}
       }
    ]
    \addplot [draw=none, fill=blue!50] coordinates {
      (KAU,11)
      (KSU, 12) 
      (KAUST,4)
      (KFUPM,1)  };
   \addplot [draw=none,fill=red!50] coordinates {
      (KAU,5)
      (KSU, 9) 
      (KAUST,2)
      (KFUPM,1)  };

    \legend{Primary HCRs, Singly Affiliated HCRs }
  \end{axis}
  \end{tikzpicture}
  \caption{The number of primary HCRs to Singly Affiliated HCRs} 
  \label{sa_sar}
  \end{figure}

We go one step further and look for the number of singly affiliated researchers for an institute. A researcher is said to be a singly affiliated if he/she has no secondary affiliation. Figure \ref{sa_sar} shows the number of singly affiliated researchers along with the total number of primary affiliated researchers. 
  \section{Conclusion}
In this work, we conducted a simple study over the ranking of Saudi Arabian universities  based on the data available on highlycited.com. Foreseeable work include analysing the universities ranking while taking into account other resources available on the web and conducting a comparative study with the 2001 list. 
\bibliographystyle{plain}
\bibliography{references}
\end{document}